# Can Large Language Model–Generated Responses Support Assessment Development? A Human-Calibrated Rasch Benchmark

Eunjeong Song · Sehee Hong
Department of Education, Korea University, Seoul, Republic of Korea

Eunjeong Song, ORCID iD: https://orcid.org/0000-0002-2302-3227
Sehee Hong, ORCID iD: https://orcid.org/0000-0001-5468-8398

## Abstract

Large language models (LLMs) are proposed as synthetic respondents for pilot testing, but their usefulness depends on whether they supply the evidence assessment development requires. We calibrated rating scale models on 14 digital-use skill items from 6,245 adults and used the human item parameters to evaluate responses generated for 1,300 demographically matched personas. LLM responses had high internal consistency (α ≈ .94) but used the lowest category in 0.2–0.3% of responses versus 13.7–21.5% for humans, and no persona chose it on every item. These gaps changed the response-category judgment on both subscales and the PC targeting judgment; on the human-calibrated scales, 12 of 14 items had infit below 0.70, indicating responses more predictable than the Rasch model expects. Preregistered changes to the prompt, category order, and persona information did not restore the human lower range. High internal consistency is insufficient evidence that LLM responses can replace human pilot data.

## Introduction

Assessment development requires evidence that response categories function as intended, that item locations suit the intended respondents, and that particular items need revision (Bond et al., 2020; Linacre, 2002; Wilson, 2005). These judgments depend on the response sample, whether the instrument is a classroom or clinical rating scale or a population survey. The *Standards for Educational and Psychological Testing* emphasize that validity evidence must be relevant to the intended population and use (American Educational Research Association,

American Psychological Association, & National Council on Measurement in Education, 2014). A sample that omits an important part of the response distribution can therefore support misleading development decisions even when conventional psychometric summaries look acceptable.

Large language models (LLMs) have been used to generate items, tests, reports, and simulated responses (Argyle et al., 2023; Hommel et al., 2022; Li et al., 2026; Lockwood et al., 2025). In this study, a *synthetic persona* is an individual profile supplied to the model, a *synthetic respondent* is the model–persona configuration, and *LLM-generated responses* are its outputs. Unlike generated items or reports, these responses become the observations on which psychometric judgments rest, so their adequacy must be judged by whether they provide the evidence that each development decision requires.

We examine three such decisions for a four-category self-report instrument. *Response-category structure* asks whether the categories are used and function as intended. *Targeting* asks whether item and step locations are appropriate for the distribution of person measures. *Item review* asks whether item fit calls for a content review. We distinguish targeting from *lower-range representation*, which asks whether the response source reproduces the lower part of the human response distribution; in this instrument, lowest-category use and all-items-lowest response patterns describe that range. In other instruments the decision-relevant range may be the upper end.

Previous studies have compared LLM and human responses at the level of item calibration and early scale development. Liu et al. (2025) compared responses to dichotomously scored algebra items under fixed human Rasch parameters and under separate calibrations and found narrower LLM proficiency distributions. Johnson et al. (2025) anchored 40 human-calibrated Likert items of a social-emotional scale and found that LLM responses to new items yielded discrimination parameters far above the human values with nearly identical thresholds across items. Säuberli et al. (2025) evaluated 18 models as pilot participants for multiple-choice items and found overconfident response distributions that temperature scaling only partly corrected. Cipriani et al. (2025) examined factor structure, measurement invariance, and score distributions and developed new scales with simulated responses before testing them in human samples; group-level structure was partly reproduced, individual-level distributions were not. Lukauskas and Šarkauskaitė (2026) audited 37 models on three Likert instruments and found over-coherent, low-variance responses that neither debiasing instructions nor few-shot conditioning removed. Lim et al. (2026) validated items with virtual respondents equipped with trait–response mediators, and Maeda (2025) aligned AI responses with human-calibrated response probabilities. Other work documents reduced diversity, directional bias, and sensitivity to response-option presentation in LLM survey responses (Bisbee et al., 2024; Dominguez-Olmedo et al., 2024; Salecha et al., 2024; Santurkar et al., 2023), and persona information limits the individual differences a model can reproduce (Hu & Collier, 2024). Demographic matching alone therefore does not establish that responses reproduce variation in the measured construct. What remains unexamined is whether the specific decisions of polytomous scale development would come out the same.

A *human-calibrated scale* is a Rasch reference scale whose item parameters are estimated from human responses and then held fixed (Rasch, 1960). It permits comparison of locations and residuals on common parameters without assuming that humans and LLMs share the construct. Cronbach's α summarizes a ratio of variances and does not test category functioning, lower-range representation, or agreement with a response model (Cronbach, 1951; Sijtsma, 2009; Thompson & Vacha-Haase, 2000).

This study makes three contributions. First, it evaluates LLM-generated responses against the specific decisions of scale development—response-category structure, targeting, and item review—rather than against global similarity indices. Second, it uses a polytomous rating scale calibrated on a national probability sample of 6,245 adults as a fixed human reference and examines the lower range of the response distribution, which demographic matching does not guarantee. Third, it preregisters and tests three generation conditions intended to restore the missing lower range. Generated responses to the same items are treated as hypothetical pilot data, so the study is a retrospective benchmark—an evaluation of generated responses against an existing human survey—rather than a prospective scale-development trial, and it does not validate the procedure for children or youth.

Research Question 1: Do human and demographically matched LLM responses support the same judgments about response-category structure, targeting, and item review?

Research Question 2: Which distributional and model-fit discrepancies on the human-calibrated scales underlie the differences, despite high internal consistency?

Research Question 3 (secondary, preregistered): Do changes to the prompt instruction, category presentation, or construct-relevant persona information reduce lower-range underrepresentation?

## Method

### Human Reference Sample and Instrument

The human reference was the public general-population microdata from the 2025 Survey on the Digital Divide (National Information Society Agency, 2026a, 2026b). The survey sampled 7,000 people aged 7 years or older nationwide by stratified probability-proportional sampling and oversampled adults aged 55 years or older. To match the persona age range, we included the 6,245 respondents aged 19 years or older. No disability-based exclusion was applied.

Seven personal computer (PC) items and seven mobile-device items asked whether respondents could perform specified tasks without help, with four categories from 1 = 전혀 그렇지 않다 (not at all) to 4 = 매우 그렇다 (very much). The construct is self-reported digital-use skill, primarily operational skill rather than objective performance (van Deursen & van Dijk, 2011; van Deursen et al., 2012). *Item difficulty* therefore denotes the self-reported skill needed to endorse a can-do item. Supplemental Table S1 gives the original wording and reference translations; subscale totals range from 7 to 28.

The study used public de-identified secondary data, synthetic personas, and generated responses. No participants were recruited or contacted, and no stakeholders took part in developing or revising the instrument. Ethical approval was not required; the rationale is stated on the title page.

### LLM-Generated Responses

We sampled 1,300 personas without replacement from NVIDIA Nemotron-Personas-Korea (Kim et al., 2026) with integer quotas proportional to the survey-weighted joint distribution of age band × education × region × sex, so that every demographic cell was represented. Persona text combined structured demographic fields and four narrative fields (Supplemental Section S3.1). This procedure matched the observed demographics, not digital-use skill within the cells.

OpenAI's open-weight gpt-oss-120b was accessed through Together AI (OpenAI, 2025; Together AI, 2026). Each request contained one persona and one original Korean item in a single user message, without a system prompt or previous item responses, so responses to different items were conditionally independent given the persona text. The model was instructed to answer from the persona's life context and to output one digit from 1 to 4. Temperature and top-p were 1.0, reasoning effort was low, and the output limit was 512 tokens; other sampling settings were provider defaults. Even at low effort the model generates an internal reasoning trace before the digit. This was the *reference condition*. Prompt text, retries, repeated generations, and provenance are documented in Supplemental Section S3.1.

**Human-Calibrated Rasch Analysis**

PC and mobile-device items were analyzed as separate seven-item content domains. Because the items have four categories, dimensionality was examined with polychoric eigenvalues within each subscale and with one- versus two-factor confirmatory factor analyses (CFAs) of the 14 items using mean- and variance-adjusted weighted least squares in Mplus 9.1 (Rhemtulla et al., 2012), and with principal component analysis of standardized residuals (DeMars, 2010; Linacre, 1998; E. V. Smith, 2002). The first residual contrast was compared with the 1.5 guideline and with its distribution under the fitted human model (Supplemental Section S1.3), treating no single cutoff as decisive (Chou & Wang, 2010). Local independence was examined with Yen's (1984) Q3 and its mean-adjusted form Q3*, with .20 as a descriptive flag (Christensen et al., 2017; Meyer, 2014). A combined 14-item model was a sensitivity analysis (Supplemental Section S5.2).

Partial credit models (PCMs; Masters, 1982) and rating scale models (RSMs; Andrich, 1978) were fitted by joint maximum likelihood estimation (JMLE) in jMetrik 4.1.1 (Meyer, 2014, 2018), unweighted because the estimator does not accept survey weights and without bias correction. Following the convention for Likert-type formats, the RSM was retained when person measures, reliability, and fit conclusions were nearly identical (Hong et al., 2005; Linacre, 2000). In the RSM, the *step parameter* for item $i$ and step $s$ is $\delta_i + \tau_s$, the item difficulty plus a threshold shared by all items; a threshold is the point, relative to item difficulty, at which two adjacent categories are equally probable. Mean item difficulty and the sum of thresholds were fixed at 0. Item order was checked against content-based expectations, with malware scanning expected to be hardest and file sending easiest (Wright & Stone, 1979). Category functioning was judged by at least 10 observations per category, average person measures increasing across categories, ordered thresholds, and category infit and outfit below 1.30 (Bond et al., 2020; Linacre, 2002).

Human item difficulties and thresholds were then fixed for scoring both sources. Full-sample person summaries used weighted likelihood estimates (WLEs), which remain finite for extreme response patterns (Magis & Verhelst, 2017; Warm, 1989). Item fit used respondents with finite maximum likelihood estimates (MLEs), as in standard Rasch software; an all-person WLE analysis was a sensitivity check. *Partial anchoring* fixed the human thresholds while re-estimating LLM item difficulties. To assess repeatability, 200 personas were generated on three occasions.

**Decision Rules and Comparison Indicators**

After examining the reference-condition results, we registered the decision rules before generating Conditions A–C. Category structure and item review used each source's own calibration, whereas targeting necessarily used the common human scale.

*Response-category structure.* Lowest-category use below 5% across items flagged sparse use for review. We compared four-category calibrations with recalibrations after merging categories 1 and 2, the remedy that sparse use ordinarily prompts, and applied the same merge to the human data to see what it would have cost, re-examining item order, fit, and person separation after recoding (Fox & Jones, 1998).

*Targeting.* A full-sample WLE mean more than 1.0 logit from the mean item difficulty of 0 flagged a mismatch; boundaries of 0.5–1.5 logits were also examined. Item-person maps and the rates of all-items-lowest and all-items-highest patterns showed which range needed attention.

*Item review.* Item fit was summarized with infit (information-weighted) and outfit (unweighted) residual mean squares, whose expected value is 1 (Wright & Masters, 1982). Infit ≥ 1.30, the registered boundary, triggered content review; because both samples exceed 1,000 respondents, the 1.1 boundary that Bond et al. (2020) tabulate for such samples from the sample-size adjustment of R. M. Smith et al. (1998) is reported alongside it. Infit below 0.70 indicates *overfit*—responses more predictable than the model expects (McNamara, 1996; Wright & Linacre, 1994)—which is not a deletion criterion (Linacre, 2012) and is unrelated to overfitting in machine learning. Infit was preferred to outfit for its lower sensitivity to isolated extreme responses, and mean squares were used rather than standardized statistics, which are uninformative at these sample sizes (Linacre, 2003).

We compared α, WLE means, standard deviations (SDs), and interquartile ranges (IQRs), extreme-pattern rates, Rasch person reliability (the proportion of person-measure variance not attributable to measurement error), and person separation (Wright & Masters, 1982). Because persona quotas followed the survey-weighted joint distribution, unweighted LLM summaries are compared with survey-weighted human summaries; item calibration and item fit were unweighted in both sources. Percentile intervals used 2,000 respondent-level bootstrap resamples with item parameters fixed (Efron & Tibshirani, 1993). As a reference for α, responses were also simulated under the human RSM from normal latent distributions matched to the human or LLM WLE mean and SD (Supplemental Section S4.1).

**Preregistration and Open Materials**

Human calibration, reference-condition comparisons, persona-content coding, and post-merge recalibration were exploratory. Before the first request for Conditions A–C, we registered removal of the no-stereotype instruction (Condition A, 400 personas), reversal of category display order (Condition B, 250 of those personas), and addition of matched human donors' access and utilization information (Condition C, the same 400 personas). The added information excluded all 14 outcome items. Registered outcomes were lowest-category use, all-items-lowest patterns, and WLE SD and IQR ratios; a registered check of three human subgroups (Supplemental Section S5.1) and departures from the plan (Supplemental Section S3.3) are also reported. Cross-validation and the Condition C diagnostic reported below were post hoc and are labeled exploratory.

The preregistration is available on OSF (https://osf.io/chx64). The preregistration is preserved unchanged. Data, materials, and code are available in Harvard Dataverse (https://doi.org/10.7910/DVN/S7QUCA). The RULER reporting checklist accompanies the manuscript, with rehabilitation-specific recommendations marked for applicability (Mallinson et al., 2022).

# Results

All 14 responses were complete for the 6,245 humans and the 1,300 reference-condition personas.

## Human Reference Scales

Within each subscale the first polychoric eigenvalue accounted for 88% and 80% of the variance and exceeded the second more than tenfold, supporting unidimensionality for Rasch analysis (DeMars, 2010); across the 14 items, a two-factor CFA fit better than a one-factor model (factor correlation = .913; Supplemental Table S2). The first standardized-residual contrasts (1.63 and 1.62, 23% of the residual variance) exceeded the 1.5 guideline and the reference obtained by re-estimating the RSM on data simulated under the human model (mean 1.21, 95% range 1.20–1.24; Supplemental Section S1.3), so some residual structure remained: in both subscales the contrast separated device-setup items from malware scanning and document writing. No raw Q3 exceeded .20; Q3* exceeded .20 for two PC pairs (.223 and .213, each pairing two hardware-setup or two software-setup tasks). Because excluding one item from each pair left the targeting and anchored-fit comparisons below unchanged, each subscale was treated as a single dimension. PCM and RSM person measures correlated above .999 with similar reliability and fit, so the RSM was retained. All four human categories had adequate frequencies, increasing average measures, ordered thresholds (PC: −3.59, 0.04, 3.55; mobile device: −2.25, −0.72, 2.97), and category infit and outfit below 1.10 (Supplemental Table S3; Figure S1). Malware scanning was the hardest and file sending the easiest item in both subscales, as expected.

## Assessment-Development Decisions (Research Question 1)

The two sources supported different judgments about response-category structure on both subscales and about targeting on the PC subscale, whereas item-review flags largely agreed (Table 1).

**Table 1**

*Assessment-Development Judgments From Human and LLM Responses*

| Judgment | Subscale | Human responses | LLM responses | Same judgment? |
|---|---|---|---|---|
| Response-category structure | PC | Lowest category used in 21.5% of responses; all categories functioning | Lowest category used in 0.18%; unused on 3 items; four-category calibration nonconvergent | No: LLM data prompt merging categories 1 and 2; human data support retaining them |
| Response-category structure | Mobile device | Lowest category used in 13.7%; all categories functioning | Lowest category used in 0.30%; unused on 2 items; extreme thresholds | No |
| Targeting | PC | WLE mean 0.31; all-items-lowest 14.2%; all-items-highest 10.7% | WLE mean 2.72; all-items-lowest 0.0%; all-items-highest 11.2% | No: LLM data call only for harder items; human data also require easier items |
| Targeting | Mobile device | WLE mean 1.21; all-items-lowest 6.5%; all-items-highest 10.8% | WLE mean 2.07; all-items-lowest 0.0%; all-items-highest 4.4% | Partly: both call for harder items; only human data show a floor |
| Item review (each source's own calibration) | Both | Infit ≥ 1.30: none. Infit ≥ 1.1: both malware items, mobile file sending | Infit ≥ 1.30: none. Infit ≥ 1.1: both malware items, mobile document writing | Yes at 1.30; largely at 1.1 |

*Note.* Means are logits on the human-calibrated scale. Human rates and means are survey weighted; LLM values are unweighted. Item review compares a human four-category calibration with an LLM three-category calibration because the LLM four-category calibrations were unstable (Supplemental Table S5). LLM = large language model; PC = personal computer; WLE = weighted likelihood estimate.

### *Response-Category Structure*

Only 16 PC and 27 mobile-device responses out of 9,100 per subscale used the lowest category. The LLM four-category PC calibration did not converge within 2,000 iterations, and the mobile-device calibration converged with thresholds of −14.7, 3.8, and 10.9. Merging categories 1 and 2 produced converged calibrations with no item at or above infit 1.30, so a developer relying on the LLM data would have merged the two lowest categories. Applying the same merge to the human data left the item order unchanged but placed 25.8% (PC) and 8.6% (mobile device) of respondents at the floor and lowered person separation from 3.31 to 2.26 and from 2.59 to 2.22 (unweighted; Supplemental Table S5). The merge would have resolved a sparsity peculiar to the LLM data at the cost of measurement in the intended population.

### *Targeting*

On PC, the human mean was near the item mean, whereas the LLM mean was 2.7 logits above it; the contrast held at boundaries from 0.5 to 1.5 logits. Human responses also contained an all-items-lowest pattern (14.2%) that was absent from the LLM responses (Figure 1). The LLM data would have called for harder items only; the human data call for easier items as well. On mobile devices both means exceeded 1.0 logit, but only the human responses showed the all-items-lowest pattern (6.5%).

### *Item Review*

At the registered 1.30 boundary, neither source flagged an item. At the 1.1 boundary, both sources flagged the two malware items (human infit 1.27 and 1.22; LLM 1.14 and 1.15), and the third flag differed (human: mobile file sending, 1.12; LLM: mobile document writing, 1.11). The LLM calibrations, however, located malware scanning 5.6 and 8.7 logits above the mean item difficulty, three to four times the human item range, so the LLM flags describe an item far outside the range of the other items rather than a within-range misfit (Supplemental Table S5).

## Distribution and Model Fit (Research Question 2)

### *Lower-Range Underrepresentation*

On the human-calibrated scales, LLM WLE means were 2.42 (PC) and 0.85 (mobile device) logits higher than the human means, with SD ratios of 0.62 and 0.73 (Table 2; Figure 1). The dispersion difference arose mainly in the tails: IQR ratios were 0.68 and 0.97. The LLM data lacked the human all-items-lowest pattern entirely and, on mobile devices, also had fewer all-items-highest patterns. In the 60-and-over, primary-education-or-less cell, 86.3% of humans but no persona showed the PC all-items-lowest pattern (Supplemental Section S4.2).

**Figure 1**

*Item-Person Map Under the Human-Calibrated Rating Scale Model*

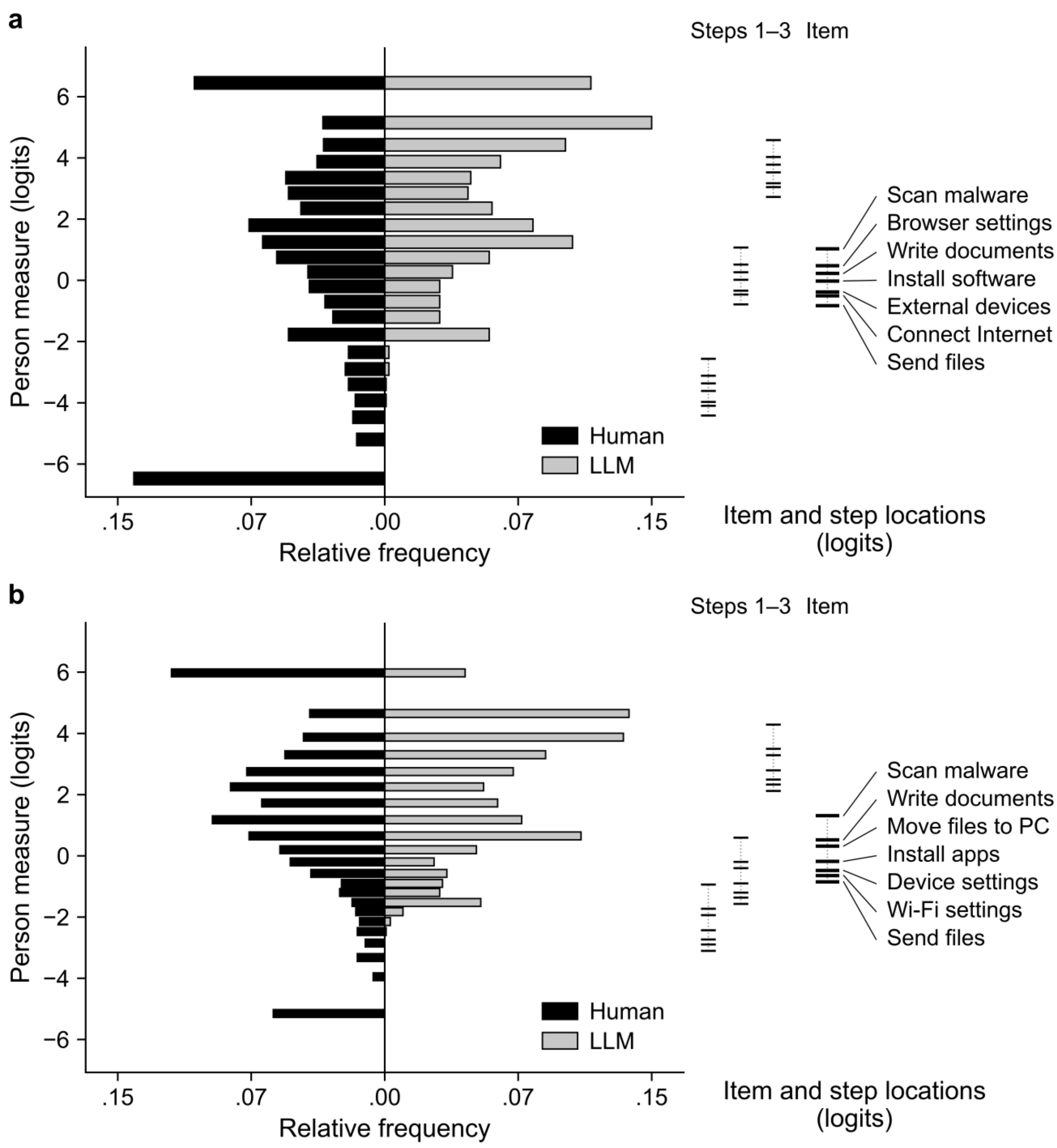

*Note.* Panels a and b show PC and mobile-device skills. Bars show full-sample WLE frequencies for humans and LLM personas; the outermost bars are extreme response patterns, not separately resolved trait values beyond the scale. Step locations equal item difficulty plus threshold. LLM = large language model; PC = personal computer; WLE = weighted likelihood estimate.

**Table 2**

*Human and LLM Responses on the Human-Calibrated Scales*

| Measure | PC human | PC LLM | PC comparison [95% CI] | Mobile human | Mobile LLM | Mobile comparison [95% CI] |
|---|---|---|---|---|---|---|
| Cronbach's α | .965 | .944 | −.021 [−.026, −.017] | .938 | .937 | −.001 [−.006, .005] |
| Mean person measure (WLE) | 0.306 | 2.721 | +2.415 [2.256, 2.584] | 1.212 | 2.065 | +0.853 [0.719, 0.994] |
| Person-measure SD (WLE) | 3.970 | 2.443 | 0.615 [0.596, 0.634] | 2.881 | 2.116 | 0.735 [0.713, 0.758] |
| Person-measure IQR (WLE) | 5.700 | 3.900 | 0.684 [0.557, 0.772] | 3.325 | 3.225 | 0.970 [0.832, 1.105] |
| Lowest-category use (%) | 21.46 | 0.18 | — | 13.66 | 0.30 | — |
| All-items-lowest pattern (%) | 14.23 | 0.00 | −14.23 [−15.10, −13.39] | 6.46 | 0.00 | −6.46 [−7.08, −5.86] |
| All-items-highest pattern (%) | 10.74 | 11.23 | +0.49 [−1.34, 2.51] | 10.81 | 4.38 | −6.43 [−7.77, −5.06] |
| Rasch person reliability | .917 | .872 | −.045 [−.053, −.039] | .871 | .855 | −.016 [−.025, −.008] |
| Rasch person separation index | 3.331 | 2.611 | 0.784 [0.756, 0.809] | 2.603 | 2.430 | 0.934 [0.901, 0.966] |

*Note.* Human summaries are survey weighted; LLM summaries are unweighted. Comparisons are LLM − human differences for α, means, extreme-pattern rates, and reliability, and LLM/human ratios for SDs, IQRs, and separation. Intervals are percentile bootstrap intervals conditional on fixed item parameters. Reliability and separation use respondents with finite MLEs; other summaries use all-person WLEs. CI = confidence interval; IQR = interquartile range; LLM = large language model; MLE = maximum likelihood estimate; PC = personal computer; SD = standard deviation; WLE = weighted likelihood estimate.

### *Overfit on the Human-Calibrated Scales*

Twelve of the 14 items had LLM infit between 0.45 and 0.68, compared with human infit of 0.80–1.27 and a mean of 0.88 in model-consistent finite-test simulations (Figure 2; Supplemental Table S4). The pattern held under cross-validated anchors and in every sensitivity analysis, although a combined 14-item scale flagged 7 rather than 12 items (Supplemental Section S5). The two malware items instead had LLM infit of 1.46 and 1.78; under partial anchoring they were 1.8 logits harder relative to the other items than in the human calibration, and their infit fell to 0.73 and 1.08. The model thus treated malware scanning as far harder than humans reported it to be, while it answered every other item more predictably than the human model expects. In the LLM's own calibration the same determinism appeared not as low infit but as a wider spread of item difficulties and thresholds (Supplemental Table S5), which is why the item-review flags in Table 1 agreed. Repeated generations of 200 personas showed within-persona variance about one third of the value implied by the human-calibrated model (ratios 0.38 and 0.34; Supplemental Section S3.1). Together with the low infit values, this indicates that the model's responses were more deterministic given the persona than human responses are given the trait.

**Figure 2**

*Lowest-Category Use and Item Infit on the Human-Calibrated Scales*

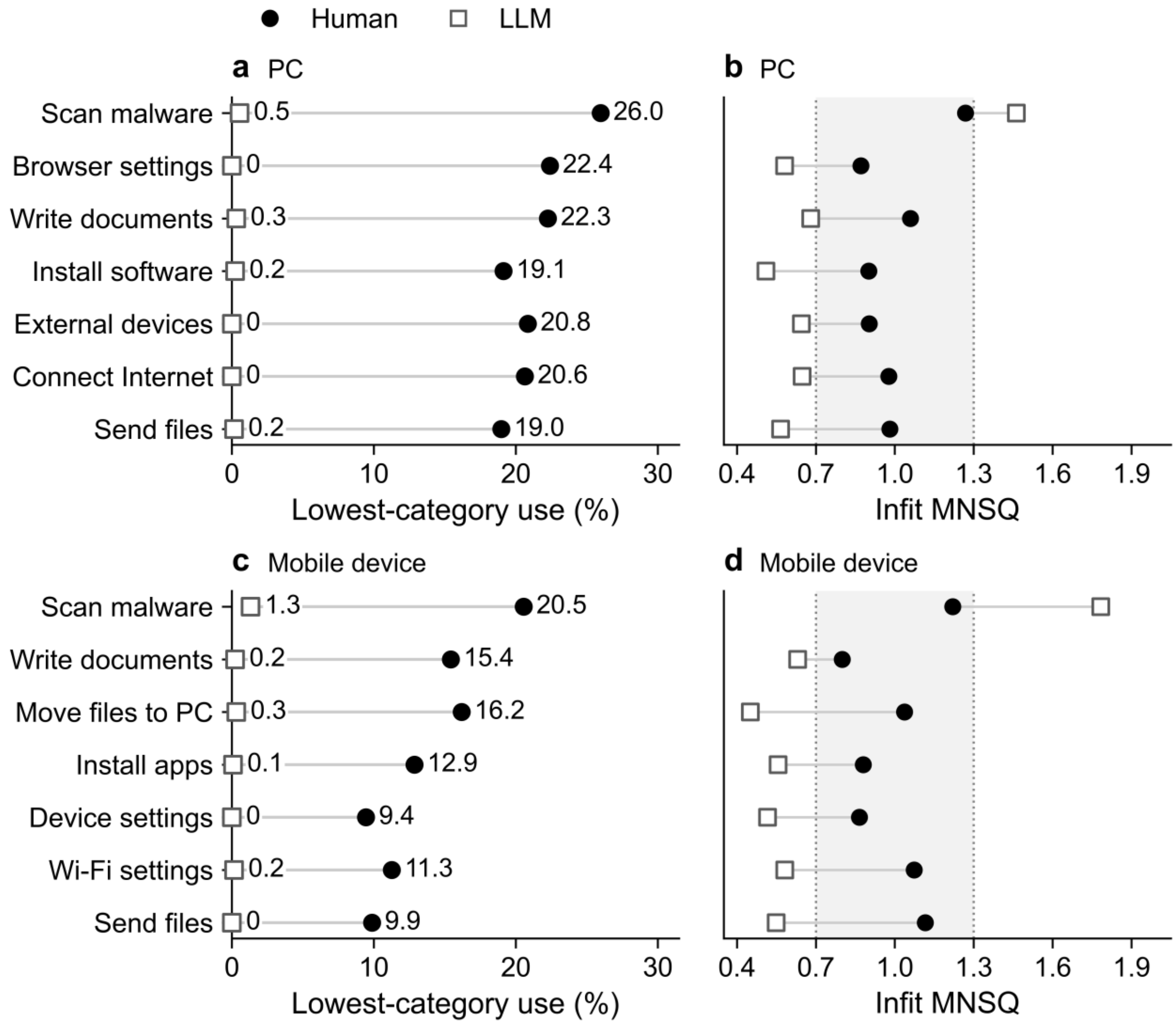

*Note.* Panels a and c show lowest-category use; panels b and d show infit among respondents with finite MLEs. Dotted lines mark 0.70 and 1.30. Items are ordered by human difficulty. LLM = large language model; MLE = maximum likelihood estimate; MNSQ = mean square; PC = personal computer.

***High Internal Consistency Despite the Discrepancies***

LLM α remained near .94, within .02 of the human values. Because α depends only on the ratio of summed item variances to total-score variance (Cronbach, 1951), and both variances were smaller in the LLM data by similar proportions (LLM/human ratios of 0.45 and 0.41 on PC and 0.51 and 0.51 on mobile devices), α barely moved while the lower range disappeared. Simulations under the human RSM with normal latent distributions matched to the LLM mean and SD produced lowest-category use of 1.81% and 3.93% and an α decrease of about .055 (Supplemental Table S9); the observed LLM data were more consistent and used the lowest category less than this model-based reference.

**Preregistered Generation Conditions (Research Question 3)**

Removing the no-stereotype instruction yielded lowest-category use of 0.21% and 0.29%, and reversing the category order yielded 0.40% and 0.34%; neither met the registered recovery or format-effect criterion. Adding construct-relevant persona information raised use to 1.25% and 1.32%, classified as partial recovery under the registered rule but still below one tenth of the human rates. No condition produced an all-items-lowest pattern, including in the two oldest, least-educated cells specified for the location check (Supplemental Table S6). WLE SD ratios relative to the human reference were 0.64–0.75 in all conditions, whereas IQR ratios ranged from 0.73 to 1.23, so the reduced dispersion again lay mainly in the tails. In the registered subgroup check, lowest-category use and low infit each distinguished the sources in five of six comparisons, so no screening indicator was superior (Supplemental Table S10).

An exploratory diagnostic of Condition C located part of the remaining deficit in the model's use of the categories rather than in missing persona information. Among the 25 personas whose added information stated that the person had not used the Internet in the past month, 17% of PC responses used the lowest category and 75% the second category, and none showed the all-items-lowest pattern; among the 374 humans who reported the same, 78% of PC responses used the lowest category and 69% showed the pattern (Supplemental Table S7). Given the same statement, the model mapped an absence of use onto "not really" rather than "not at all." The tested changes therefore did not reproduce the human lower range, and the deficit persisted even when the persona carried the relevant information.

## Discussion

Despite internal consistency as high as that of the human data, the LLM responses did not supply the evidence that category and targeting decisions require. They scarcely used the lowest category, contained no all-items-lowest pattern, and would have led a developer to merge categories and add only harder items—decisions that the human data contradict. Item-review flags largely agreed, and on the human-calibrated scales 12 of 14 items were overfit.

### Implications for Assessment Development

Category frequencies describe the response source, not whether a category is needed in the intended population. Here the merge that the LLM data prompted would have nearly doubled the human all-items-lowest rate on PC and reduced person separation. Developers should keep the original category frequencies, treat merging as a substantive revision rather than a numerical remedy, and re-examine item order, fit, and separation after recoding (Fox & Jones, 1998; Linacre, 2002).

Targeting judgments likewise require the relevant human variation. Demographic quotas reproduced the observed demographics but not the low-skill respondents within them, and augmenting personas with construct-relevant information recovered only a small part of the lower range. Part of the human floor consists of adults who do not use a PC at all; whether a scale should cover them with easier items or with a gate item is a content decision, but only the human data raise it. For other instruments the missing respondents may occupy the upper range, and extension to children, youth, or screening decisions requires evidence from those populations and uses.

Systematically low infit across multiple items on the human-calibrated scale identified responses more predictable than the human model expects. Together with the low within-persona variance across repeated generations and the Condition C diagnostic, the most consistent explanation is that the model answered nearly deterministically given the persona, that the demographic personas carried little information about skill within a cell, and that even an explicit statement of non-use was mapped onto the second category rather than the first. Temperature and top-p of 1.0 rule out truncation of the sampling distribution as the cause, but the reasoning trace that precedes the digit may itself narrow the answer distribution, and temperatures above 1.0 or logit-level calibration (Säuberli et al., 2025) were not tested; prompt, model, and serving effects were not manipulated jointly. Low infit did not outperform lowest-category use in the registered subgroup check, so it is a discrepancy signal on a reference scale rather than a detector of synthetic data. The same reasoning explains why α stayed high: it summarizes variance ratios, which reduced dispersion and reduced error left nearly unchanged (Sijtsma, 2009).

Four checks follow for developers who use generated responses:

1. Report the original category frequencies.
2. Report the rates of all-items-lowest and all-items-highest patterns.
3. When human data for the instrument or a related one exist, examine targeting and fit on the human-calibrated scale.
4. Do not revise response categories or population coverage on the basis of generated responses alone.

Generated responses may still support workflow checks and explicitly hypothetical sensitivity analyses.

## Limitations and Future Directions

The study covers one Korean self-report instrument, one model, one persona source, and item-by-item generation, and it does not establish performance for objective tests or other age groups. Construct-relevant persona information was limited: the Condition C template listed the devices a person had rather than stating the absence of a computer, and the augmentation used human variables that a genuine pilot might not have. The instruction to choose a category even with insufficient information was retained in all conditions, the reasoning step that precedes each answer was not manipulated, and public item wording may have appeared in model training. The human RSMs were useful references rather than exact or population-invariant models: the two-factor CFA had an RMSEA of .097 with excellent incremental fit, a pattern expected when loadings are high and residual dependence is small but *n* is large; differential item functioning across human groups was not examined; JMLE parameters were not bias corrected; and the bootstrap intervals omit calibration and survey-design uncertainty. Human test–retest reliability was unavailable, and repeated generation does not substitute for it. Future work should vary instruments, persona information, and generation architectures and examine direct alignment with human responses (Lim et al., 2026; Maeda, 2025).

High internal consistency did not show that the examined procedure preserved the evidence that category and targeting decisions require. Category use, extreme-pattern rates, and fit on a human-calibrated scale revealed what α could neither establish nor rule out.

## Statements and Declarations

### Ethical Considerations

Ethical approval was not required. This study used only publicly available, de-identified secondary data, synthetic personas, and LLM-generated responses. The authors did not recruit or interact with participants and had no access to personally identifiable information. Such research does not constitute human subjects research as defined in Article 2, subparagraph 1 of the *Bioethics and Safety Act* of the Republic of Korea and Article 2(1) of its Enforcement Rule; even if so construed, it would be exempt from Institutional Review Board review under Article 15(2) of the Act and Article 13(1)3 of the Enforcement Rule as research using existing, publicly available data. Accordingly, no committee reviewed this study and no approval number exists.

### Consent to Participate

Not applicable.

### Consent for Publication

Not applicable.

### Declaration of Conflicting Interests

The authors declared no potential conflicts of interest with respect to the research, authorship, and/or publication of this article.

### Funding

The authors received no financial support for the research, authorship, and/or publication of this article.

### AI-Use Disclosure

DeepL was used for translation and English editing and OpenAI Codex for checking the analysis code; the authors reviewed all outputs and take full responsibility for the manuscript and code. The model studied, gpt-oss-120b, generated the synthetic responses described in the Method.

# Supplemental Material

This supplement provides the item wording, reference-model checks, source-specific recalibrations, generation protocol, preregistered outcomes, an exploratory diagnostic of Condition C, and robustness checks referenced in the main text. Complete output is in the analysis package. LLM = large language model; MLE = maximum likelihood estimate; PC = personal computer; PCM = partial credit model; RSM = rating scale model; WLE = weighted likelihood estimate.

## S1. Items and Human Reference Models

### *S1.1 Items and Factor Structure*

All items used four categories: 1 = 전혀 그렇지 않다 (not at all), 2 = 그렇지 않은 편이다 (not really), 3 = 그런 편이다 (somewhat), and 4 = 매우 그렇다 (very much). Table S1 preserves the Korean wording administered in the source survey and used for generation; the English translations are for reference only.

**Table S1**

*Items Used in the Analysis*

(a) PC. Stem: 귀하는 PC(데스크탑/노트북)를 통해 다음의 활동을 스스로 얼마나 하실 수 있습니까? [To what extent can you do the following activities on your own using a PC (desktop/laptop)?]

| Variable | Short name | Item as administered (Korean) | English translation |
|---|---|---|---|
| Q6_6 | Scan malware | 나는 PC 의 악성코드(바이러스, 스파이웨어 등)를 검사/치료할 수 있다. | I can scan for and remove malware (viruses, spyware, etc.) on a PC. |
| Q6_3 | Browser settings | 나는 웹 브라우저(크롬, 사파리, 엣지, 웨일 등)에서 내가 원하는 환경을 설정할 수 있다(팝업창 차단, 텍스트 크기 설정, 보안 및 시작 홈페이지 설정 등). | I can configure a web browser (Chrome, Safari, Edge, Whale, etc.) the way I want (blocking pop-ups, setting the text size, security and start-page settings, etc.). |

| Variable | Short name | Item as administered (Korean) | English translation |
|---|---|---|---|
| Q6_7 | Write documents | 나는 PC 에서 문서(한글, 엑셀, 파워포인트 등) 등을 작성할 수 있다. | I can create documents (HWP, Excel, PowerPoint, etc.) on a PC. |
| Q6_1 | Install software | 나는 필요한 프로그램(소프트웨어)을 컴퓨터에 설치/삭제/업데이트 할 수 있다. | I can install, remove, or update the programs (software) I need on a computer. |
| Q6_4 | External devices | 나는 PC 에 다양한 외장기기(디지털 카메라, 프린터, 스캐너, USB 외장하드 등)를 연결하여 이용할 수 있다. | I can connect and use various external devices (digital camera, printer, scanner, USB external hard drive, etc.) with a PC. |
| Q6_2 | Connect Internet | 나는 PC 에 유선 또는 무선 인터넷을 스스로 연결해서 사용할 수 있다(단, IP 설정은 제외). | I can connect a PC to wired or wireless Internet on my own and use it (excluding IP configuration). |
| Q6_5 | Send files | 나는 PC 에 있는 파일을 인터넷을 통해 다른 사람에게 전송할 수 있다. | I can send files on a PC to other people over the Internet. |

(b) Mobile device. Stem: 귀하는 스마트기기(스마트폰, 스마트패드 등)를 통해 다음의 활동을 스스로 얼마나 하실 수 있습니까? [To what extent can you do the following activities on your own using a smart device (smartphone, smart pad, etc.)?]

| Variable | Short name | Item as administered (Korean) | English translation |
|---|---|---|---|
| Q7_6 | Scan malware | 나는 스마트기기의 악성코드(바이러스, 스파이웨어 등)를 검사/치료할 수 있다. | I can scan for and remove malware (viruses, spyware, etc.) on a smart device. |
| Q7_7 | Write documents | 나는 스마트기기에서 문서(메모장, 워드)를 작성할 수 있다. | I can create documents (notepad, word processor) on a smart device. |
| Q7_3 | Move files to PC | 나는 스마트기기에 있는 파일을 PC 로 옮길 수 있다. | I can move files on a smart device to a PC. |
| Q7_5 | Install apps | 나는 앱을 스마트기기에 설치/삭제/업데이트할 수 있다. | I can install, remove, or update apps on a smart device. |
| Q7_1 | Device settings | 나는 스마트기기에서 디스플레이/소리/보안/알림/입력방법 등의 환경설정을 할 수 있다. | I can configure settings such as display, sound, security, notifications, and input method on a smart device. |

| Variable | Short name | Item as administered (Korean) | English translation |
|---|---|---|---|
| Q7_2 | Wi-Fi settings | 나는 스마트기기에서 무선 랜(와이파이, 기가와이파이 포함) 설정을 할 수 있다. | I can configure wireless LAN (including Wi-Fi and giga Wi-Fi) on a smart device. |
| Q7_4 | Send files | 나는 스마트기기에 있는 파일/사진 등을 다른 사람에게 전송할 수 있다. | I can send files or photos on a smart device to other people. |

*Note.* Items are ordered from harder to easier within each subscale using the human estimates in Table S3. Item mapping and response direction were checked against the National Information Society Agency (2026a) report; survey-weighted positive-response rates reproduced the published item rates to one decimal place.

**Table S2**

*Factor-Structure Checks and Comparison of the Human Rasch Models*

(a) Fourteen-item CFA models (Mplus 9.1, WLSMV).

| Model | $\chi^2$ | *df* | RMSEA [90% CI] | CFI | TLI | SRMR |
|---|---|---|---|---|---|---|
| One factor | 8406.70 | 77 | .132 [.129, .134] | .984 | .981 | .039 |
| Two correlated factors (PC and mobile device) | 4512.67 | 76 | .097 [.094, .099] | .992 | .990 | .023 |

(b) One-factor models within subscales (lavaan 0.6-17, WLSMV) and polychoric eigenvalues.

| Subscale | $\chi^2$ | *df* | RMSEA [90% CI] | CFI | TLI | SRMR | Standardized loadings | Eigenvalues (1st, 2nd, 3rd) |
|---|---|---|---|---|---|---|---|---|
| PC | 1029.08 | 14 | .108 [.102, .113] | .998 | .996 | .014 | .890–.945 | 6.18, 0.23, 0.16 |
| Mobile device | 762.35 | 14 | .093 [.087, .098] | .995 | .993 | .023 | .819–.908 | 5.61, 0.41, 0.24 |

(c) PCM and RSM results (jMetrik 4.1.1).

| Subscale | Model | Person reliability | Person separation | Infit median [min, max] | Outfit median [min, max] | Items with infit or outfit ≥ 1.30 | Items with infit or outfit < 0.70 |
|---|---|---|---|---|---|---|---|
| PC | PCM | .916 | 3.300 | 0.94 [0.87, 1.27] | 0.93 [0.85, 1.27] | 0 | 0 |
| PC | RSM | .916 | 3.295 | 0.98 [0.87, 1.27] | 0.94 [0.86, 1.27] | 0 | 0 |
| Mobile device | PCM | .871 | 2.593 | 0.96 [0.84, 1.26] | 0.95 [0.81, 1.32] | 1 | 0 |

| Subscale | Model | Person reliability | Person separation | Infit median [min, max] | Outfit median [min, max] | Items with infit or outfit ≥ 1.30 | Items with infit or outfit < 0.70 |
|---|---|---|---|---|---|---|---|
| Mobile device | RSM | .869 | 2.578 | 1.04 [0.80, 1.22] | 1.04 [0.79, 1.31] | 1 | 0 |

*Note.* Human $N = 6{,}245$; all calibrations were unweighted. Every CFA $\chi^2$ had $p < .001$. In the 14-item two-factor model the factors correlated .913 and standardized loadings were .842–.945; the loadings in panel (b) are from the separate one-factor models. lavaan used the theta parameterization (Muthén & Muthén, 2017; Rosseel, 2012). The scaled difference test, computed in lavaan rather than by subtracting WLSMV $\chi^2$ values, favored the two-factor model, $\Delta\chi^2(1) = 928.28$, $p < .001$ (Satorra, 2000). RMSEA is elevated in models with few degrees of freedom and high loadings (Kenny et al., 2015); with $n = 6{,}245$, the small residual dependence reported in Section S1.3 is enough to raise it while the incremental indices remain excellent. PCM–RSM person-measure correlations were .99997 (PC) and .99995 (mobile device). Reliability and separation here are unweighted calibration summaries; survey-weighted values appear in main-text Table 2. Item reliability was ≥ .997 and item separation ≥ 19.4 for all four Rasch models. CFA = confirmatory factor analysis; CFI = comparative fit index; CI = confidence interval; RMSEA = root mean square error of approximation; SRMR = standardized root mean square residual; TLI = Tucker–Lewis index; WLSMV = mean- and variance-adjusted weighted least squares.

### *S1.2 Human RSM Parameters*

**Table S3**

*Human RSM Parameters and Category Diagnostics*

(a) Item parameters and fit.

| Subscale | Item | Difficulty | *SE* | Infit | Outfit |
|---|---|---|---|---|---|
| PC | Scan malware | 1.03 | 0.03 | 1.27 | 1.27 |
| PC | Browser settings | 0.48 | 0.03 | 0.87 | 0.86 |
| PC | Write documents | 0.23 | 0.03 | 1.06 | 1.05 |
| PC | Install software | −0.02 | 0.03 | 0.90 | 0.92 |
| PC | External devices | −0.38 | 0.03 | 0.90 | 0.88 |
| PC | Connect Internet | −0.50 | 0.03 | 0.98 | 0.96 |
| PC | Send files | −0.83 | 0.03 | 0.98 | 0.94 |
| Mobile device | Scan malware | 1.31 | 0.02 | 1.22 | 1.31 |
| Mobile device | Write documents | 0.52 | 0.02 | 0.80 | 0.79 |
| Mobile device | Move files to PC | 0.32 | 0.03 | 1.04 | 1.04 |
| Mobile device | Install apps | −0.18 | 0.03 | 0.88 | 0.84 |
| Mobile device | Device settings | −0.48 | 0.03 | 0.87 | 0.84 |

| Subscale | Item | Difficulty | *SE* | Infit | Outfit |
|---|---|---|---|---|---|
| Mobile device | Wi-Fi settings | −0.64 | 0.03 | 1.07 | 1.04 |
| Mobile device | Send files | −0.85 | 0.03 | 1.12 | 1.05 |

(b) Thresholds and category fit.

| Subscale | Threshold | Estimate | *SE* | Category infit | Category outfit |
|---|---|---|---|---|---|
| PC | $\tau_1$ | −3.59 | 0.03 | 1.02 | 1.03 |
| PC | $\tau_2$ | 0.04 | 0.02 | 0.95 | 0.92 |
| PC | $\tau_3$ | 3.55 | 0.02 | 0.95 | 0.95 |
| Mobile device | $\tau_1$ | −2.25 | 0.03 | 1.08 | 1.07 |
| Mobile device | $\tau_2$ | −0.72 | 0.02 | 1.05 | 0.98 |
| Mobile device | $\tau_3$ | 2.97 | 0.02 | 0.95 | 0.95 |

(c) Category frequencies and average person measures.

| Subscale | Category | Observed frequency | Minimum item frequency | Average person measure |
|---|---|---|---|---|
| PC | 1 | 8,949 | 1,133 | −5.46 |
| PC | 2 | 9,697 | 1,070 | −0.87 |
| PC | 3 | 14,087 | 1,778 | 1.60 |
| PC | 4 | 10,982 | 1,161 | 4.77 |
| Mobile device | 1 | 5,612 | 553 | −3.54 |
| Mobile device | 2 | 6,222 | 580 | −0.23 |
| Mobile device | 3 | 17,925 | 2,300 | 1.17 |
| Mobile device | 4 | 13,956 | 1,256 | 4.07 |

*Note.* jMetrik JMLE estimates from the unweighted human sample, with mean item difficulty and the sum of thresholds fixed at 0 within each subscale (Meyer, 2014, 2018); no bias correction was applied. Item fit uses respondents with finite MLEs; average measures are full-sample WLEs. The step parameter for item $i$ and step $s$ is $\delta_i + \tau_s$. As a scale sensitivity, an approximate $(L - 1)/L = 6/7$ parameter shrinkage (Wright & Douglas, 1977) reduced the human–LLM WLE mean differences from 2.42 to 2.16 logits (PC) and from 0.85 to 0.77 (mobile device) without changing the targeting or anchored-fit classifications (`table-jmle_bias_sensitivity.csv`). *SE* = standard error.

***S1.3 Category Curves, Residual Structure, and Person Fit***

**Figure S1**

*Category Probability Curves of the Human RSMs*

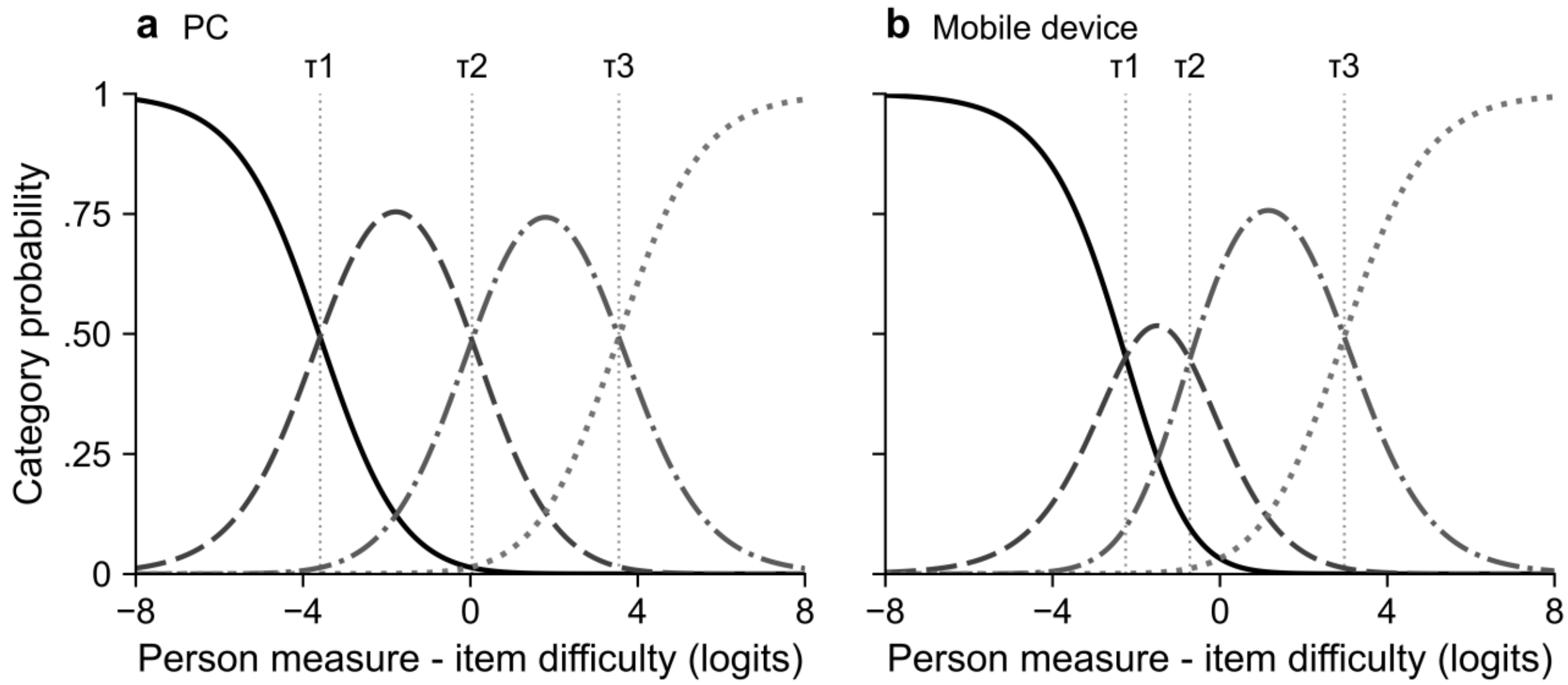

*Note.* Panels a and b show PC and mobile-device skills. The horizontal axis is person measure minus item difficulty in logits. All items within a subscale share the same thresholds and curves; vertical dotted lines mark $\tau_1$–$\tau_3$ from Table S3. Each category is the most probable over a distinct interval.

The first standardized-residual contrasts had eigenvalues of 1.63 and 1.62, about 23% of the seven-item residual variance (Linacre, 1998; E. V. Smith, 2002). To obtain a reference for a seven-item scale, responses were generated 100 times under the human RSM from the human-matched normal latent distribution of Section S4.1 ($n$ = 6,245), the RSM was re-estimated by uncorrected JMLE on each replicate, and the first eigenvalue of the correlation matrix of standardized residuals among respondents with finite MLEs was recorded. The package estimator reproduced the jMetrik values for the observed data (1.63 and 1.62). Under the model the first eigenvalue averaged 1.21 ($SD$ = 0.01; 2.5th–97.5th percentiles 1.20–1.24 on PC and 1.20–1.23 on mobile devices), and no replicate reached 1.5 (`table-first_contrast_reference.csv`). The observed contrasts therefore reflect residual structure beyond the model, although no single cutoff was treated as decisive (Chou & Wang, 2010). On PC, software installation, Internet connection, browser settings, and external devices loaded positively and file sending, malware scanning, and document writing negatively; on mobile devices, device settings, Wi-Fi settings, app installation, and file sending loaded positively and moving files to a PC, malware scanning, and document writing negatively.

Raw Q3 had negative off-diagonal means (−.163 and −.164) and no pair above .20 (Yen, 1984). Mean-adjusted Q3* exceeded .20 for two PC pairs, Internet connection–external devices (.223) and software installation–browser settings (.213); the mobile-device maximum was .134 (Christensen et al., 2017; Meyer, 2014). Recalibrating without one item from each flagged pair left the targeting and anchored-fit results unchanged (`table-local_dependence_sensitivity.csv`). Person infit below 0.70 occurred in 38% (human) and 43% (LLM) of PC records and in 37% and 53% of mobile-device records; no person was removed for fit.

## S2. Anchored Fit and Separate Recalibrations

### *S2.1 Human-Calibrated Fit and Partial Anchoring*

**Table S4**

*Anchored Item Fit and Partial-Anchoring Item Positions*

| Subscale | Item | Human infit / outfit | LLM infit / outfit | Partial-anchoring LLM difficulty | LLM − human difficulty |
|---|---|---|---|---|---|
| PC | Scan malware | 1.27 / 1.27 | 1.46 / 1.49 | 2.80 | 1.77 |
| PC | Browser settings | 0.87 / 0.86 | 0.58 / 0.56 | −0.07 | −0.55 |
| PC | Write documents | 1.06 / 1.05 | 0.68 / 0.64 | −0.49 | −0.72 |
| PC | Install software | 0.90 / 0.92 | 0.51 / 0.49 | 0.12 | 0.15 |
| PC | External devices | 0.90 / 0.88 | 0.64 / 0.58 | −0.27 | 0.11 |
| PC | Connect Internet | 0.98 / 0.96 | 0.65 / 0.56 | −1.25 | −0.75 |
| PC | Send files | 0.98 / 0.94 | 0.57 / 0.50 | −0.84 | −0.01 |
| Mobile device | Scan malware | 1.22 / 1.31 | 1.78 / 2.11 | 3.05 | 1.73 |
| Mobile device | Write documents | 0.80 / 0.79 | 0.63 / 0.67 | −0.27 | −0.80 |

| Subscale | Item | Human infit / outfit | LLM infit / outfit | Partial-anchoring LLM difficulty | LLM − human difficulty |
|---|---|---|---|---|---|
| Mobile device | Move files to PC | 1.04 / 1.04 | 0.45 / 0.47 | −0.11 | −0.43 |
| Mobile device | Install apps | 0.88 / 0.84 | 0.56 / 0.55 | −1.04 | −0.86 |
| Mobile device | Device settings | 0.87 / 0.84 | 0.52 / 0.53 | −0.02 | 0.46 |
| Mobile device | Wi-Fi settings | 1.07 / 1.04 | 0.58 / 0.53 | −0.36 | 0.28 |
| Mobile device | Send files | 1.12 / 1.05 | 0.55 / 0.49 | −1.25 | −0.40 |

*Note.* Fit statistics use unweighted respondents with finite MLEs: human $n$ = 4,745 (PC) and 5,141 (mobile device); LLM $n$ = 1,154 and 1,243. Partial anchoring fixes the human thresholds and estimates LLM difficulties with mean item difficulty 0 (standard errors 0.05–0.07 logits); human difficulties are in Table S3. These positions are descriptive, not a formal test of differential item functioning.

Six of seven LLM items per subscale had infit below 0.70; including extreme patterns with WLE scoring left the same six items below 0.70. A model-based finite-test comparison generated 200 response matrices at the finite-MLE LLM locations under the human RSM: mean item infit was 0.88, with minimum values of 0.76 and 0.74 and no item below 0.70. Under partial anchoring, malware infit decreased from 1.46 to 0.73 (PC) and from 1.78 to 1.08 (mobile device) while the other six items remained below 0.70; LLM item-difficulty ranges were 4.05 versus 1.86 logits (PC) and 4.30 versus 2.17 (mobile device), with rank correlations of .79 and .68.

### *S2.2 Source-Specific Calibration and Category Merging*

**Table S5**

*Separate Calibrations With Four Categories and After Merging Categories 1 and 2*

| Subscale | Source | Categories | Lowest-category count (minimum per item) | Convergence | Thresholds | Item-difficulty range | Items with infit ≥ 1.30 / ≥ 1.1 | Person reliability | Person separation | All-items-lowest (%) |
|---|---|---|---|---|---|---|---|---|---|---|
| PC | Human | 4 | 8,949 (1,133) | Converged (43) | −3.62, 0.04, 3.58 | 1.87 | 0 / 1 (malware 1.28) | .916 | 3.31 | 13.7 |
| PC | Human | 3 | — | Converged (24) | −1.83, 1.83 | 2.06 | 0 / 1 (malware 1.29) | .836 | 2.26 | 25.8 |

| Subscale | Source | Categories | Lowest-category count (minimum per item) | Convergence | Thresholds | Item-difficulty range | Items with infit ≥ 1.30 / ≥ 1.1 | Person reliability | Person separation | All-items-lowest (%) |
|---|---|---|---|---|---|---|---|---|---|---|
| PC | LLM | 4 | 16 (0) | Not converged (2,000-iteration limit) | Not interpreted | — | — | — | — | 0.0 |
| PC | LLM | 3 | — | Converged (65) | −3.17, 3.17 | 7.95 | 0 / 1 (malware 1.14) | .872 | 2.61 | 6.3 |
| Mobile device | Human | 4 | 5,612 (553) | Converged (32) | −2.27, −0.72, 2.99 | 2.18 | 0 / 2 (malware 1.23; send files 1.12) | .870 | 2.59 | 6.1 |
| Mobile device | Human | 3 | — | Converged (25) | −1.88, 1.88 | 2.42 | 1 / 1 (malware 1.38) | .832 | 2.22 | 8.6 |
| Mobile device | LLM | 4 | 27 (0) | Converged (135), extreme thresholds | −14.73, 3.81, 10.91 | Not interpreted | — | — | — | 0.0 |
| Mobile device | LLM | 3 | — | Converged (103) | −3.63, 3.63 | 11.82 | 0 / 2 (malware 1.15; write documents 1.11) | .850 | 2.38 | 6.5 |

*Note.* All calibrations use the package's Python JMLE on each source's own data, without bias correction, matching the jMetrik reference calibration; extreme response patterns are excluded from estimation and fit, and reliability and separation use respondents with finite MLEs. Human four-category thresholds reproduce jMetrik within .03 logits. All-items-lowest rates are unweighted, so the human four-category values differ from the survey-weighted values in main-text Table 2. Positions are not linked across calibrations; item-difficulty ranges are not equated differences.

Merging categories 1 and 2 produced numerical convergence in the LLM data but did not recover the lower range: after merging, the LLM all-items-lowest rate reflects personas answering 2 on every item (6.3% and 6.5%), against 25.8% and 8.6% for humans. In the human data the merge left the item-difficulty order of Table S3 unchanged in both subscales. At the registered 1.30 boundary, no usable calibration flagged an item except that merging the human mobile-device categories raised malware infit to 1.38. At 1.1, both sources flagged the malware items; the human four-category calibration additionally flagged mobile file sending and the LLM three-category calibration mobile document writing. The LLM difficulties for malware scanning were 5.63 and 8.73 logits, three to four times the human item range, so the LLM calibrations describe malware scanning as an outlying item rather than as within-range misfit; in the LLM's own calibration the determinism that appears as low infit on the human-calibrated scale is absorbed into wider item and threshold spreads. Merged human calibrations lowered person separation from 3.31 to 2.26 (PC) and from 2.59 to 2.22 (mobile device); a revised category structure must be re-examined after recoding rather than accepted on convergence alone (Fox & Jones, 1998).

## S3. Generation Protocol and Registered Outcomes

### *S3.1 Personas, Prompt, Settings, and Repeated Generations*

Integer demographic quotas used the largest-remainder method. Each persona contained age, sex, education, occupation, province, family type, housing type, and marital status, followed by basic, professional, family, and cultural narratives. Requests did not share earlier item responses. Exploratory keyword coding, completed before registration, used 62 digital-device/service terms and 10 low-capability expressions: digital and low-capability mentions appeared in 28.8% and 5.8% of personas, and in the 60-and-over, primary-education-or-less cell, where 86.3% of human respondents had the PC all-items-lowest pattern, only 1 of 59 personas mentioned low capability.

The original Korean prompt was:

```
당신은 아래 인물이다. 인물의 생활맥락과 경험에 근거하여 답하라.
인구집단에 대한 고정관념이나 일반적인 한국인의 성향을 추측하지 말라.
정보가 충분하지 않더라도 아래 인물의 구체적인 경험만을 근거로 가장 가까운
범주를 선택하라.

[인물]
{페르소나 텍스트}

[질문]
{문두}
```

```
"{문항}"

[응답범주]
1 = 전혀 그렇지 않다
2 = 그렇지 않은 편이다
3 = 그런 편이다
4 = 매우 그렇다

반드시 1, 2, 3, 4 중 하나의 숫자만 출력하라. 설명, 문장부호, JSON을 추가하지 마라.
```

Reference translation:

```
You are the person below. Answer on the basis of this person's life context and experience.
Do not guess from stereotypes about demographic groups or from general tendencies of Koreans.
Even if the information is insufficient, choose the closest category based only on this person's concrete experience.

[Person]
{persona text}

[Question]
{stem}
"{item}"

[Response categories]
1 = not at all
2 = not really
3 = somewhat
4 = very much

Output exactly one digit among 1, 2, 3, and 4. Do not add explanations, punctuation, or JSON.
```

The model identifier was `openai/gpt-oss-120b` at Together AI's serverless endpoint, with temperature and top-p of 1.0, low reasoning effort, and a 512-token output limit; the model returns a reasoning trace before the digit even at low effort. An output-format failure allowed one retry and network or server errors up to four. Request seeds were deterministic per

persona–item cell (base 20260805), but the hosted service does not guarantee identical regeneration, so the stored responses are the canonical analysis data. Across the 23,800 reference and repeat response cells (1,300 + 200 × 2 personas × 14 items), 26 failed attempts were retried and no cell remained missing; per-call logs for the reference condition were not preserved, so those failures cannot be classified. Conditions A, C, and B had 5, 1, and 0 format failures, each resolved by one retry, and their logs are archived.

For 200 personas generated on three occasions, the WLE intraclass correlation, ICC(C,1), was .958 (PC) and .954 (mobile device; McGraw & Wong, 1996). Within-persona variances of 0.220 and 0.188 $\text{logit}^2$ were compared with the model-implied conditional variances of 0.581 and 0.556 for the 157 and 185 personas with finite MLEs on every occasion; the ratios were 0.38 [0.32, 0.44] and 0.34 [0.28, 0.40] over 2,000 persona resamples. This is a descriptive repeatability comparison, not a human test–retest study.

### *S3.2 Preregistered Generation Conditions and an Exploratory Diagnostic*

Condition A removed the no-stereotype instruction; Condition B displayed the categories in reverse order without changing the number–label pairs; Condition C added access and utilization information from a demographically matched human donor assigned without replacement within the same age × education × region × sex cell, excluding all 14 outcome items. The Condition C sentences stated the devices present in the home, whether the Internet had been used in the past month, the number of days of smart-device use, and the main purposes of use; the absence of a computer was therefore implicit in the device list rather than stated. Conditions A and C used the same 400 personas and B a 250-persona subset. In the preregistered disclosure, adding the donor variables to demographic-cell indicators increased the explained human total-score variance from .561 to .635 (PC) and from .547 to .661 (mobile device; $n$ = 6,245).

**Table S6**

*Preregistered Generation Outcomes*

(a) Category use and all-items-lowest patterns.

| Condition | Subscale | *n* | Lowest-category use, % [95% CI] | All-items-lowest, count/*n* |
|---|---|---|---|---|
| A: Prompt instruction | PC | 400 | 0.21 [0.00, 0.54] | 0/400 |
| A: Prompt instruction | Mobile device | 400 | 0.29 [0.07, 0.54] | 0/400 |
| C: Persona information | PC | 400 | 1.25 [0.68, 1.93] | 0/400 |
| C: Persona information | Mobile device | 400 | 1.32 [0.71, 2.07] | 0/400 |
| B: Category order | PC | 250 | 0.40 [0.17, 0.69] | 0/250 |
| B: Category order | Mobile device | 250 | 0.34 [0.11, 0.63] | 0/250 |

(b) Paired contrasts and dispersion relative to the human reference.

| Condition | Subscale | Paired difference, pp [95% CI] | WLE SD ratio [95% CI] | WLE IQR ratio [95% CI] |
|---|---|---|---|---|
| A: Prompt instruction | PC | +0.071 [−0.071, 0.286] | 0.641 [0.609, 0.671] | 0.728 [0.645, 0.939] |
| A: Prompt instruction | Mobile device | +0.143 [−0.071, 0.357] | 0.750 [0.714, 0.787] | 1.226 [0.903, 1.381] |
| C: Persona information | PC | +1.107 [0.536, 1.786] | 0.690 [0.657, 0.721] | 0.899 [0.728, 0.987] |
| C: Persona information | Mobile device | +1.179 [0.571, 1.894] | 0.646 [0.603, 0.687] | 0.797 [0.548, 0.932] |
| B: Category order | PC | +0.229 [−0.171, 0.571] | 0.649 [0.607, 0.686] | 0.811 [0.711, 0.899] |
| B: Category order | Mobile device | +0.171 [−0.114, 0.457] | 0.745 [0.696, 0.792] | 1.053 [0.800, 1.381] |

*Note.* The full reference condition ($n$ = 1,300) had lowest-category use of 0.18% (PC) and 0.30% (mobile device) and no all-items-lowest pattern; the matched reference personas had 0.143% (A/C) and 0.171% (B) on each subscale, and paired differences compare each condition with those same personas. Human survey-weighted rates were 21.46% and 13.66%, with all-items-lowest rates of 14.23% and 6.46%. Dispersion ratios use full-sample WLEs with survey-weighted human and unweighted generated responses. Intervals are 2,000 percentile bootstrap resamples with fixed item parameters; condition and reference personas were resampled as pairs. All-zero counts are reported as counts because resampling an all-zero indicator yields a degenerate interval. CI = confidence interval; IQR = interquartile range; pp = percentage points; SD = standard deviation.

The registered recovery criterion for lowest-category use was at least 50% of the human rate (PC ≥ 10.75%; mobile device ≥ 6.85%); rates below 1.0% were nonrecovery, and intervening values were partial recovery. The format-effect rule for B required at least three times the reference rate and at least a 1-percentage-point increase. Condition C met partial recovery only; A and B met neither criterion. No condition produced an all-items-lowest pattern, including in the two older, lower-education cells specified for C's location check.

To locate the remaining deficit, an exploratory post hoc diagnostic (registration §4.4) compared category use under two of the added statements with the human respondents whose survey variables defined those statements (Table S7). Given the statement that the person had not used the Internet in the past month, the model chose the second category for three quarters of the PC and mobile-device responses and the lowest category for about one fifth, and no persona showed the all-items-lowest pattern, whereas the humans who reported the same chose the lowest category for 78% and 69% of their responses and 69% and 55% showed the pattern. When the device list omitted a computer, the lowest category was again rare (2.8% and 2.3%) against 34% and 22% for humans. Human values are unweighted; the human groups are larger and were not matched on other characteristics, so the comparison is descriptive.

**Table S7**

*Exploratory Condition C Diagnostic: Category Use Given the Same Access or Use Statement*

| Statement | Subscale | Source | Persons | Category 1 (%) | Category 2 (%) | Category 3 (%) | Category 4 (%) | All-items-lowest, count (%) |
|---|---|---|---|---|---|---|---|---|
| No Internet use in the past month | PC | Human | 374 | 78.0 | 12.6 | 7.1 | 2.3 | 259 (69.3) |
| No Internet use in the past month | PC | LLM, Condition C | 25 | 16.6 | 74.9 | 5.7 | 2.9 | 0 (0.0) |
| No Internet use in the past month | Mobile device | Human | 374 | 69.3 | 12.0 | 15.0 | 3.7 | 205 (54.8) |
| No Internet use in the past month | Mobile device | LLM, Condition C | 25 | 20.6 | 73.1 | 4.0 | 2.3 | 0 (0.0) |
| No computer at home | PC | Human | 2,213 | 34.3 | 24.4 | 25.1 | 16.2 | 564 (25.5) |
| No computer at home | PC | LLM, Condition C | 136 | 2.8 | 62.3 | 22.0 | 12.9 | 0 (0.0) |

| Statement | Subscale | Source | Persons | Category 1 (%) | Category 2 (%) | Category 3 (%) | Category 4 (%) | All-items-lowest, count (%) |
|---|---|---|---|---|---|---|---|---|
| No computer at home | Mobile device | Human | 2,213 | 22.1 | 17.7 | 37.9 | 22.4 | 236 (10.7) |
| No computer at home | Mobile device | LLM, Condition C | 136 | 2.3 | 32.8 | 30.5 | 34.5 | 0 (0.0) |

*Note.* Percentages are of item responses (7 per person per subscale). For the LLM, statements are identified from the rendered context sentences of Condition C; for humans, from the source-survey variables that generated those sentences (Internet use in the past month; desktop or laptop ownership). This analysis was not preregistered (`table-condition_c_context_diagnostic.csv`).

### *S3.3 Registration and Reporting Departures*

The reference analyses had already been examined when the study was registered; registration preceded the first Condition A–C request on 2026-08-19 at 05:22 UTC, and the archived registration is unchanged.

**Table S8**

*Registered Plan, Actual Conduct, and Reporting Clarifications*

| Plan or reporting issue | Actual conduct and reason | Consequence for interpretation |
|---|---|---|
| Threshold-axis category fit in the registered category decision | Discontinued because it is not a standard category-fit statistic; frequencies, average measures, threshold order, and category fit were retained | Category conclusions rest on the reported diagnostics |
| Generic variance-compression simulation as the α reference | Replaced by simulations under the human RSM with matched WLE moments; the registered simulation remains archived | Simulations are model-specific references, not causal decompositions |
| Joint recovered/not-recovered interpretation of A and C | C showed partial recovery, a branch not specified in the joint table | No unique remaining cause is inferred from the registered conditions |
| Screening hierarchy claimed if low infit distinguished ≥ 5 of 6 subgroup comparisons | Both low infit and lowest-category use distinguished 5 of 6 | No hierarchy is claimed |
| Full-sample WLE framework for fit | Primary item fit uses finite MLEs; full-sample WLE fit is a sensitivity | Extreme patterns remain in distribution summaries but not in the primary fit denominator |
| Registered 1.30 item-review boundary only | The 1.1 boundary for samples above 1,000 (R. M. Smith et al., 1998) is reported alongside | Item review is compared at two boundaries |
| Additional unregistered analyses | Cross-validation, the parameter-shrinkage sensitivity, the first-contrast reference simulation, and the Condition C diagnostic are reported as post hoc exploratory checks | They are not confirmatory validation |

## S4. Reliability and Distributional Comparisons

### *S4.1 Model-Based Reference Simulations for α*

Because α depends only on the ratio of summed item variances to total-score variance (Cronbach, 1951; Lord & Novick, 1968) and Rasch person reliability only on the ratio of mean error variance to person-measure variance (Wright & Masters, 1982), neither statistic can show why both variances fell. As a model-based reference, responses were generated under the human RSM from normal latent distributions in three conditions: human-matched (reference $n$ = 6,245), location shift only, and location and dispersion matched (generated $n$ = 1,300), with generating means and SDs chosen to reproduce the corresponding full-sample WLE moments. Each condition used 500 replications; Monte Carlo standard errors were below .001 for ratios and differences and below .06 percentage points for rates.

**Table S9**

*Model-Based Reference Simulations and Observed Comparisons*

(a) PC.

| Measure | Human-matched | Location shift only | Location and dispersion matched | Observed LLM |
|---|---|---|---|---|
| Full-sample WLE mean | 0.28 | 2.71 | 2.71 | 2.72 |
| Full-sample WLE SD ratio | — | 0.898 | 0.615 | 0.615 |
| Lowest-category use (%, item mean) | 21.14 | 7.65 | 1.81 | 0.18 |
| All-items-lowest pattern (%) | 8.90 | 2.28 | 0.06 | 0.00 |
| Mean item infit (finite MLE) | 0.88 | 0.89 | 0.89 | 0.73 |
| Items with infit < 0.70 | 0 | 0 | 0 | 6 |
| Δα | — | −0.003 | −0.055 | −0.021 |
| ΔRasch person reliability | — | −0.010 | −0.065 | −0.045 |

(b) Mobile device.

| Measure | Human-matched | Location shift only | Location and dispersion matched | Observed LLM |
|---|---|---|---|---|
| Full-sample WLE mean | 1.19 | 2.05 | 2.05 | 2.07 |
| Full-sample WLE SD ratio | — | 0.964 | 0.735 | 0.735 |
| Lowest-category use (%, item mean) | 13.76 | 8.44 | 3.93 | 0.30 |
| All-items-lowest pattern (%) | 3.04 | 1.47 | 0.14 | 0.00 |
| Mean item infit (finite MLE) | 0.88 | 0.88 | 0.88 | 0.72 |
| Items with infit < 0.70 | 0 | 0 | 0 | 6 |
| Δα | — | −0.004 | −0.056 | −0.001 |
| ΔRasch person reliability | — | −0.013 | −0.063 | −0.016 |

*Note.* Simulated ratios and differences are relative to the human-matched simulation in the same replication; observed comparisons use survey-weighted human and unweighted LLM data, and fit uses respondents with finite MLEs. The human-matched simulation reproduced lowest-category use but not the all-items-lowest cluster (8.90% vs. observed 14.23% on PC; 3.04% vs. 6.46% on mobile devices), so the normal latent family does not fully reproduce the human distribution.

### *S4.2 Demographic-Cell Descriptions*

Among the eight age × education cells with at least 30 records in both sources, LLM unweighted seven-item totals exceeded human totals in 15 of 16 subscale comparisons; in the 60-and-over, primary-education-or-less cell, PC means were 7.48 (human) and 14.63 (LLM), and 86.3% of humans but no persona showed the all-items-lowest pattern. Both between-cell and within-cell total-score variances were smaller in the LLM data (ratios .430/.394 on PC and .578/.448 on mobile devices; `table-cell_variance_decomposition.csv`).

## S5. Robustness Checks

### *S5.1 Registered Subgroup Check*

The registered check used three mutually exclusive human subgroups not previously examined for this purpose. Lowest-category use and all-items-lowest rates distinguished the sources if the human rate was at least three times the LLM rate and at least 1 percentage point higher; low infit distinguished them if at most one quarter of human items and at least half of LLM items had infit below 0.70.

**Table S10**

*Registered Human-Subgroup Results*

| Subscale | Human subgroup | *n* | Lowest category (%) | Result | All-items-lowest (%) | Result | Human items with infit < 0.70 | Result |
|---|---|---|---|---|---|---|---|---|
| PC | Aged 19–39, upper secondary | 517 | 2.18 | Distinguishes | 0.97 | Does not | 0 / 7 | Distinguishes |
| PC | Aged 40–59, tertiary | 1,296 | 4.10 | Distinguishes | 1.31 | Distinguishes | 0 / 7 | Distinguishes |
| PC | Aged 60+, tertiary | 178 | 14.77 | Distinguishes | 6.18 | Distinguishes | 2 / 7 | Does not |
| Mobile device | Aged 19–39, upper secondary | 517 | 0.97 | Does not | 0.19 | Does not | 1 / 7 | Distinguishes |
| Mobile device | Aged 40–59, tertiary | 1,296 | 1.37 | Distinguishes | 0.15 | Does not | 0 / 7 | Distinguishes |
| Mobile device | Aged 60+, tertiary | 178 | 8.67 | Distinguishes | 1.69 | Distinguishes | 0 / 7 | Distinguishes |

*Note.* Comparators were the full reference LLM data (lowest-category use 0.18%/0.30%, six low-infit items per subscale, no all-items-lowest pattern). The human calibration included the evaluated subgroups, so this is not an out-of-sample check. Human malware infit reached 1.49 and 1.37 in two comparisons.

Lowest-category use and low infit each distinguished five of six comparisons and the all-items-lowest indicator three, so no indicator was shown superior. Recalibrating without each evaluated subgroup (six converged models) left the five-of-six result unchanged.

### *S5.2 Fourteen-Item Scale*

A post hoc single RSM for all 14 items converged with thresholds of −2.146, −0.492, and 2.638. Unweighted human versus LLM summaries were: lowest-category use 16.65% versus 0.24%; all-items-lowest 5.60% versus 0.00%; WLE mean 0.77 versus 2.07 logits; SD 2.912 versus 2.040; and Rasch person reliability .944 versus .916. Survey-weighted human α was .970 and unweighted LLM α .967. Items with infit below 0.70 numbered 0 of 14 versus 7 of 14, and items at or above 1.30 numbered 1 of 14 in each source. The direction of the distributional and anchored-fit patterns therefore remained, although the low-infit count fell from 12 to 7 (`table-single_scale_robustness.csv`).

### *S5.3 Cross-Validated Calibration*

This post hoc check used five-fold assignment stratified by age × education × region × sex; each four-fold calibration scored the held-out human fold and the full LLM sample with the training-fold parameters, and all 10 models converged. Mean human infit in the held-out folds was 1.001 (range 0.953–1.020) and 1.005 (0.963–1.060), with no held-out human item below 0.70; mean LLM infit was 0.732 (0.723–0.736) and 0.728 (0.722–0.735), with six low-infit items per subscale in every fold.

## Supplemental References

## RULER Checklist

Guideline: Rasch Reporting Guideline for Rehabilitation Research (RULER; Mallinson et al., 2022, *Archives of Physical Medicine and Rehabilitation, 103*(7), 1477–1486, Table 2; EQUATOR Network). Codes, levels, and numbering follow Table 2; the recommendation text is condensed. Levels: R = Required, S = Strongly recommended, E = Encouraged, N = Not encouraged. Responses: Y = reported, P = partly reported, N = not reported, NA = not applicable. "MS" = main text; "SM" = Supplemental Material. Locations refer to the final manuscript sections and tables; page numbers are added after journal formatting.

Study type in RULER terms. A rating scale model was applied to an existing public self-report instrument (14 items, two 7-item subscales) with a human calibration sample ($n$ = 6,245), and a synthetic response source (1,300 LLM-generated personas) was evaluated on the human-calibrated scale and in its own calibration. The instrument was not revised and no item or person was removed; the category-merging analyses are exploratory comparisons of what a revision would have cost, not substantive revisions. Status codes describe reporting, not proof of validity.

### 1. Conceptual/content validation methods

| Code | Level | Recommendation (condensed) | Response | Location | Notes |
|---|---|---|---|---|---|
| 1.1 | R | Content validity evaluation and stakeholder role in item development | NA | MS Method, *Human Reference Sample and Instrument* | Existing national-survey items (National Information Society Agency, 2026a); the authors did not develop or revise items. The construct (self-reported operational digital skill; van Deursen & van Dijk, 2011) and item content are stated; the original survey's content-validation procedures are outside the study. |
| 1.2 | R | Threshold map or key form | Y | MS Figure 1; SM Table S3; SM Figure S1 | A raw-score-to-measure key form is not provided (see 4.2.4). |
| 1.3 | R | Figures labeled for the upper and lower ranges of persons and items | Y | MS Figure 1 and note; MS Table 2 | |
| 1.4 | R | Item calibration tables ordered hierarchically | Y | SM Table S3(a); SM Table S4 | |

| Code | Level | Recommendation (condensed) | Response | Location | Notes |
|---|---|---|---|---|---|
| 1.5 | R | Rating scale definitions | Y | MS Method, *Human Reference Sample and Instrument*; SM Table S1 | |
| 1.6 | R | New assessments: conceptual model and item/category selection | NA | — | Existing assessment. |
| 1.7 | R | Existing assessments: prior hierarchies and expected item order | Y | MS Method, *Human-Calibrated Rasch Analysis*; MS Results, *Human Reference Scales*; SM S1.2 | Content-based hardest/easiest expectations and observed order are reported. |
| 1.8 | R | Discussion of whether results support or challenge the trait definition | Y | MS Discussion, *Implications* and *Limitations* | |

## 2. Structural validation methods

### *2.1 Rating scale structure*

| Code | Level | Recommendation (condensed) | Response | Location | Notes |
|---|---|---|---|---|---|
| 2.1.1 | R | Type of rating scale model | Y | MS Method, *Human-Calibrated Rasch Analysis*; MS Results, *Human Reference Scales*; SM Table S2 | |
| 2.1.2 | R | Categories with < 10 responses; counts per category by item | P | SM Table S3(c); MS Figure 2; SM Table S5; package category tables | Aggregate/minimum frequencies and lowest-category item rates are reported; exhaustive category-by-item counts are in the package. |
| 2.1.3 | R | Monotonicity of thresholds | Y | MS Results, *Human Reference Scales*; SM Table S3(b); SM Figure S1 | |
| 2.1.4 | R | Implications of revising categories/items | Y | MS Results, *Response-Category Structure*; MS Discussion, *Implications*; SM Table S5; SM S2.2 | The cost of merging categories 1 and 2 in the human data (all-items-lowest rate, person separation, item fit) is reported; item order after recoding was checked and was unchanged. |

***2.2 Unidimensionality, PCAR, item fit, person fit***

| Code | Level | Recommendation (condensed) | Response | Location | Notes |
|---|---|---|---|---|---|
| 2.2.5 | R | Overall item–trait interaction $\chi^2$ | NA | SM Table S2; SM S1.3; SM S2.1 | The item–trait interaction test is a RUMM statistic; jMetrik (JMLE) does not produce it. Overall model–data fit is evaluated through item infit/outfit, person fit, standardized-residual principal components, residual correlations, and the CFA structure. |
| 2.2.6 | R | Item fit table, values used, and rationale | Y | MS Method, *Decision Rules and Comparison Indicators*; MS Results, *Item Review*; SM Tables S4–S5 | Mean squares at the registered 1.30 boundary and the 1.1 boundary recommended for samples above 1,000 (Smith et al., 1998); standardized statistics are not used as criteria. |
| 2.2.7 | R | PCAR criteria and rationale | Y | MS Results, *Human Reference Scales*; SM Table S2(b); SM S1.3 | Residual eigenvalues are distinguished from total test variance; the first contrast is compared with the 1.5 guideline and with its simulated distribution under the fitted RSM (`table-first_contrast_reference.csv`). |
| 2.2.8 | R | Reasons for item misfit, implications, resolution | P | MS Results, *Overfit on the Human-Calibrated Scales*; SM S2.1–S2.2 | Fit and partial-anchoring sensitivity are reported; the malware items' displacement is interpreted; no item was deleted. |
| 2.2.9 | R | Effect on person measures of removing/retaining misfitting items | P | SM S1.3 (item-exclusion sensitivity); SM S5.2 | Exclusion and analysis-unit sensitivities are described; a full person-by-person change table is not reproduced. |
| 2.2.10 | R | Percentage of misfitting persons | P | SM S1.3; package `table-person_fit.csv` | Proportions with person infit below 0.70 are reported; full person-fit output remains in the package. |
| 2.2.11 | R | Reasons for person misfit, implications, resolution | P | SM S1.3 | No person was removed for fit, and causes of person misfit were not established. |
| 2.2.12 | R | Reintroduction of removed persons after anchoring; item displacement | NA | MS Method; SM S2.1 | No persons were removed for fit. Exclusion of extreme patterns from finite-MLE fit calculations is distinguished from deletion from the dataset. |
| 2.2.a | S | Winsteps: Smith iterative approach | NA | — | jMetrik was used, not Winsteps. |

| Code | Level | Recommendation (condensed) | Response | Location | Notes |
|---|---|---|---|---|---|
| 2.2.b | S | First-contrast eigenvalue < 2 and variance share < 10% | P | MS Results, *Human Reference Scales*; SM S1.3 | First-contrast eigenvalues (1.63, 1.62) are below 2 but above the model-based reference (1.21); the contrast content is described; the Winsteps percentage of total variance is unavailable. |

### *2.3 Measurement accuracy*

| Code | Level | Recommendation (condensed) | Response | Location | Notes |
|---|---|---|---|---|---|
| 2.3.13 | R | PSR without extreme persons; PSI and strata (Winsteps) | Y | MS Table 2; SM Table S2(c); SM Table S5 | Strata are not reported (Winsteps-specific); they can be derived from the separation index. |
| 2.3.14 | R | Wright's sample-independent PSR adjustment | N | SM Table S3 note | Not applied; the (L − 1)/L scale sensitivity does not replace it. |
| 2.3.15 | R | Floor and ceiling effects > 15% | Y | MS Tables 1–2; MS Results, *Targeting*; MS Figure 1; SM Table S5 | Reported although below 15% in the four-category scale, because all-items-lowest patterns are central to the argument; the merged human scale exceeds 15% on PC. |
| 2.3.16 | R | Mean ± SD of person measures; persons vs. items | Y | MS Table 2; MS Results, *Targeting* | |
| 2.3.a | S | Interpretation of PSR thresholds | N | MS Table 2 | Values are reported; decision-use thresholds are not applied because the research question concerns population coverage, not the instrument's use for individual decisions. |
| 2.3.b | E | RUMM users: PSI and strata without extremes | NA | — | Not RUMM. |
| 2.3.c | N | If Cronbach's α is reported, state records included | Y | MS Table 2; MS Results, *High Internal Consistency Despite the Discrepancies*; MS Introduction | α is reported deliberately because its behavior under synthetic data is a research question; all records were included (*n* = 6,245 survey weighted; *n* = 1,300). |

### *2.4 Iterations of analysis*

| Code | Level | Recommendation (condensed) | Response | Location | Notes |
|---|---|---|---|---|---|
| 2.4.17 | R | Iterations table | Y | SM Tables S2, S5, and S8 | Model comparisons, category merging, and departures are reported; no substantive item revision was made. |

### 3. External validation methods

| Code | Level | Recommendation (condensed) | Response | Location | Notes |
|---|---|---|---|---|---|
| 3.1.a | S | Visual presentation of associations across the continuum | NA | — | No external criterion instrument was available; human–LLM comparisons are not external validity coefficients. |
| 3.1.b | E | A priori hypotheses about association strengths | NA | SM S3.2 | No convergent/discriminant measures; the preregistered hypotheses concern category use and dispersion under the generation conditions. The Condition C diagnostic (SM Table S7) is exploratory. |

### 4. Consequential validation methods

| Code | Level | Recommendation (condensed) | Response | Location | Notes |
|---|---|---|---|---|---|
| 4.1.1 | R | Conditional MDC for person-level change | NA | — | No change measurement; single administration. |
| 4.1.2 | R | Effect sizes from Rasch measures for group change | NA | MS Table 2 | No change over time; between-source differences are expressed in logits with bootstrap intervals. |
| 4.1.a | E | Relative precision ratio | NA | — | No comparison instrument. |
| 4.2.3 | R | Time points and settings | Y | MS Method; SM S3.1–S3.3 | The human survey year and generation chronology/settings (including the reasoning trace preceding each answer) are reported; per-call logs for the reference condition are unavailable. |
| 4.2.4 | R | Score-to-measure table and key form use | NA | SM Table S3; analysis package | The study does not propose the instrument for clinical decisions; fixed parameters and open scoring code allow conversion. |
| 4.2.5 | R | Measurement error and stability of measures | Y | MS Table 2; SM Table S3 note; SM S2.2; SM S3.1 | All calibrations of observed data are uncorrected JMLE; the $(L - 1)/L$ scale sensitivity left targeting and anchored-fit classifications |

| Code | Level | Recommendation (condensed) | Response | Location | Notes |
|---|---|---|---|---|---|
| | | | | | unchanged; repeated generations are reported. |
| 4.2.6 | R | Cautionary language for small or pilot samples | Y | MS Discussion, *Implications* and *Limitations* | |
| 4.2.a | E | Conversion tables | N | SM Table S3 | Not provided; parameters and code are public. |
| 4.3.7 | R | Stakeholder involvement | Y | MS Method, *Human Reference Sample and Instrument* | No stakeholders were involved; this is stated explicitly. |
| 4.3.8 | R | Future directions across settings, ages, sex, ethnicity, impairments | Y | MS Discussion, *Implications* and *Limitations and Future Directions* | Impairment-specific applicability is not addressed (self-report population survey). |

**5. Measurement invariance, reproducibility, and reliability**

| Code | Level | Recommendation (condensed) | Response | Location | Notes |
|---|---|---|---|---|---|
| 5.1.1 | R | Residual correlations ≥ .2 above average | Y | MS Results, *Human Reference Scales*; SM S1.3 | Raw Q3 and mean-adjusted Q3* are reported; all pairs are in the package (`table-jmetrik_q3_pairs.csv`). |
| 5.1.2 | R | Consequences of item revision on person measures (Luppescu) | NA | — | No item was revised. |
| 5.1.3 | R | State when no consequential LID was found | P | MS Results, *Human Reference Scales*; SM S1.3 | Two PC Q3* flags and a first residual contrast above the model-based reference were found; exclusion sensitivity is reported; absence of local dependence is not claimed. |
| 5.1.4 | R | Method for response dependency and results | Y | MS Method, *Human-Calibrated Rasch Analysis*; SM S1.3 | |
| 5.1.5 | R | Repeated measures: one time point; hierarchy stability | NA | SM S3.1 | Human data are single-administration; repeated generations describe repeatability of the model, not human hierarchy stability. |

| Code | Level | Recommendation (condensed) | Response | Location | Notes |
|---|---|---|---|---|---|
| 5.1.6 | R | Rationale for DIF grouping variables | N | MS Limitations | No demographic DIF analysis was performed; between-source invariance is not established. |
| 5.1.7 | R | Data amendments after DIF | NA | — | No DIF-based amendments were made. |
| 5.2.a | S | Reliability across time, raters, forms | P | SM S3.1; MS Limitations | Repeated LLM generations are reported; human test–retest, interrater, and alternate-form reliability were not assessed. |
| 5.2.b | S | Limitations for reliability analyses not conducted | Y | MS Discussion, *Limitations* | |
| 5.2.c | E | Graphical association of Rasch measures (Bland–Altman) | NA | MS Figure 1 | No second measure of the same persons; distributions are compared graphically instead. |

## 6. Practical application and clinical implementation

| Code | Level | Recommendation (condensed) | Response | Location | Notes |
|---|---|---|---|---|---|
| 6.1.1 | R | Exclusion of persons with disabilities | Y | MS Method, *Human Reference Sample and Instrument* | No disability-based exclusion was applied; this does not establish disability representativeness of the source survey. |
| 6.1.2 | R | Scoring sheet, manual, permanent link | P | SM Tables S1 and S3; title page Data Availability; package | Items, parameters, and scoring code are available; a clinical scoring manual is not claimed; permanent DOIs replace the anonymized links on acceptance. |
| 6.2.3 | R | Access, cost, and guidance | P | Title page Data Availability; package README and LICENSE | Source access and reuse restrictions are described. |
| 6.2.a | E | Links to assessment databases | NA | — | Not a rehabilitation assessment. |
| 6.3.4 | R | Reading level and inclusive wording | NA | SM Table S1 | The instrument was not developed or revised; items are reproduced verbatim. |
| 6.3.a | E | Language and version used | Y | MS Method, *LLM-Generated Responses*; SM Table S1 | |

| Code | Level | Recommendation (condensed) | Response | Location | Notes |
|---|---|---|---|---|---|
| 6.3.b | E | Availability in other languages | Y | SM Table S1 | The instrument exists in Korean; the English translations are for readers only. |

**Summary**

| Level | Items | Y | P | N | NA |
|---|---|---|---|---|---|
| Required | 44 | 23 | 8 | 2 | 11 |
| Strongly recommended | 6 | 1 | 2 | 1 | 2 |
| Encouraged | 8 | 2 | 0 | 1 | 5 |
| Not encouraged | 1 | 1 | 0 | 0 | 0 |
| Total | 59 | 27 | 10 | 4 | 18 |

Required recommendations marked P, N, or NA are accompanied by reasons above. Reporting status is not a certification that all measurement properties have been established.